# SPATIAL AND LUMINOSITY PROPERTIES OF THE PERSEUS-PISCES REDSHIFT SURVEY

Francesco Sylos Labini[1,2], Marco Montuori [2,3] Luciano Pietronero[2]

April, 25, 1995

([1]) Dipartimento di Fisica, Università di Bologna, Italy

([2]) Dipartimento di Fisica, Università di Roma "La Sapienza"
P.le A. Moro 2, I-00185 Roma, Italy.

([3]) Dipartimento di Fisica, Università di Cosenza

## Abstract

We analyze the spatial and the luminosity properties of the Perseus-Pisces redshift survey. We find that the two point correlation function (CF) $\Gamma(r)$ is a power law up to the sample effective depth ($\sim 30h^{-1}Mpc$), showing the fractal nature of the galaxy distribution in this catalog. The fractal dimension turns out to be $D \approx 2$, somewhat larger than the value obtained for the CfA1 catalog. We also consider the CF $\xi(r)$ and in particular the behavior of the "*correlation length*" $r_0$ ($\xi(r_0) \equiv 1$) as function of the sample size. In this respect we find, unambiguously, that the luminosity segregation effect is not supported by any experimental evidence. In addition we have studied the number-distance $(N(r))$ and number-counts $(N(m))$ relations in the VL subsamples finding a good agreement with the properties of a fractal distribution. In particular our conclusion is that the $N(r)$ relation permits to extend the analysis of the fractal nature up to a deeper depth than that reached by the CF analysis, and, we find evidence for fractal properties up to the limiting depth of $140 - 150h^{-1}Mpc$. We stress that the $N(m)$ relation must be studied in VL subsamples rather than in magnitude limited ones as usually done. In this latter case the Malmiquist bias affects the behavior of $N(m)$, while in VL samples we find that $\alpha = D/5$ as predicted for fractal distributions, with $D \approx 2$. Hence we find a perfect agreement among the $\Gamma(r)$, $N(r)$ and $N(m)$ analyses. Finally we have considered the correlations between galaxy positions and luminosities by means of the multifractal analysis. We find clear evidence for self-similar behavior of the whole luminosity-space distribution. We stress that all these analyses are not based on any " a priori" assumptions and are consistent each other. These results confirm and extend those of Coleman & Pietronero (1992).

Subject headings: galaxies: clustering – galaxies: distances and redshifts– methods:numerical

# 1 Introduction

The distribution of galaxies in space have been investigated very intensively in the last years. Several recent galaxy redshift surveys such as CfA 1 (Huchra et al., 1983), CfA 2 (De Lapparent et al., 1986; da Costa et al., 1994; Park et al., 1994), SSRS1 (Da Costa et al., 1988), SSRS2 (Da Costa et al., 1994), Perseus Pisces (Haynes & Giovanelli, 1988), pencil beams surveys (Broadhurst et al., 1990) and ESP (Vettolani et al., 1994), have uncovered remarkable structures such as filaments, sheets, superclusters and voids. These galaxy catalogues probe scales from $\sim 100 - 200 h^{-1} Mpc$ for the wide angle surveys, up to $\sim 1000 h^{-1} Mpc$ for the deeper pencil beam surveys, and show that the Large-Scale Structures (LSS) are the characteristic features of the visible matter distribution. One of the most important issues raised by these catalogues is that the scale of *the largest inhomogeneities* is comparable with *the extent of the surveys* so that the largest known structures are limited only by *the boundaries of the surveys* in which they are detected. Hence from these data a new picture emerges in which the scale of homogeneity seems to shift to a very large value, not still identified. Sometimes this problem is underestimated by the argument that LSS have small amplitudes (Peebles et al., 1991). However this argument is misleading because the amplitude have no physical meaning in self-similar distributions (Baryshev et al., 1994),

The usual correlation function analysis performed by the $\xi(r)$, leads to the identification of the "correlation length" $r_0 \approx 5 h^{-1} Mpc$ (Davis & Peebles, 1983). This result appears incompatible with the existence of LSS of order of $50 - 200 h^{-1} Mpc$. In fact, according to this result, the distribution should become smooth and regular at distances larger than $r_0$, while it is evident that this is not the case. The main problem of the $\xi(r)$-analysis is that it is based on the *assumption* that the distribution of galaxies in the available samples is homogenous. We refer the reader to Coleman & Pietronero (1992 - hereafter CP92) for a detailed discussion of this subject (see also Borgani, 1995).

The basic idea is to perform a correlation analysis that does not require any a priori assumption (Pietronero, 1987, Coleman et al., 1988). This new correlation analysis reconciles the statistical studies with the observed LSS. The result is that the CfA1 redshift survey for galaxies and the Abell catalog for clusters, have fractal properties up to their effective depths. Here we perform the same kind of analysis on the Perseus-Pisces redshift surveys (Haynes & Giovanelli, 1988) that is deeper than CfA1 and has a better statistics. Moreover we also perform various other tests to investigate the effect of the eventual correlation between spatial and luminosity distributions.

Several authors claimed that the available redshift surveys do not contain a *fair sample* of the Universe. For example also the combined sample CfA2 and SSRS2 have been declared to be "$not fair$" (Da Costa et al., 1994). One should actually distinguish between a "statistical fair sample", which is a sample where there are enough points to derive some statistical properties unambiguously, and a *homogeneous* sample. Homogeneity is a property that can be present or not, but that has nothing to do with the statistical validity of the sample. CP92 showed that a small sample like CfA 1 is statistically fair up to a distance that can be defined unambiguously. Nevertheless they found that this sample is not homogenous, but that it has fractal nature up to $\approx 20 h^{-1} Mpc$. Above this distance the sample looses its statistical meaning.

Hence one of the main important tasks of observational cosmology is the identification of the homogeneity scale $\lambda_0$ above which the distribution of galaxies may really become homogenous. Baryshev et al. (1994) considered this test as a crucial one in order to discriminate among alternatives cosmological models.

In *section 2* we describe the catalog and how we have built the volume limited (VL) subsamples. In *section 3* we briefly introduce the basic properties of fractal distributions and in *section 4* we present the results of the correlation function (CF) $\Gamma(r)$ analysis. We perform such an analysis following CP92, without using any weighting scheme or treatment of boundary conditions. The problem of the weighting schemes is a very crucial one because various authors (Guzzo et al., 1991; Provenzale et al., 1994) find different results from ours only in the region in which they apply this treatment to the data. We discuss this point in the following, but we emphasize that our results are in perfect agreement with those of Guzzo et. al. (1991) in the region where they do not make any corrections to the data. In this situation it is likely that the presumed homogeneity is actually an artifact of weighting schemes. We will show that this is actually the case.

In *section 5* we study the standard CF $\xi(r)$ and in particular we analyze the sample depth dependence of the so-called "correlation length" $r_0$ defined by $\xi(r_0) \equiv 1$. We find that $r_0$ scales linearly with the sample size in agreement with the result of CP92 for CfA1. Various authors have discussed this property as an effect of the luminosity segregation phenomenon (Da Costa et al., 1988; Park et al., 1994). In *section 6* we test this hypothesis showing that it is not supported by any experimental result. On the contrary we show that the fractal nature of the galaxy distribution naturally describes the behavior of $r_0$ as a function of the sample size (see also Sylos Labini & Amendola, 1995, Amendola & Sylos Labini, 1995)

In *section 7* we study the classical relations that are the number-distance $N(r)$ and the number-count $N(m)$ in VL subsamples. We discuss why these relations have to be studied only in such kind of samples if one wants to avoid the effect of the Malmquist bias. We stress that, for example, the number-counts relation is always studied in magnitude limited catalogs and in such samples it is seriously affected by selection effects. On the contrary, we show that only in the VL samples one can recover the genuine properties of the number-counts, and only there one should study this problem avoiding any bias or selection effect. The $N(r)$ relation permits us to extend the study on the fractal nature of the sample up to much larger length scales than those allowed by the CF analysis ($\approx 140 - 150\ h^{-1}Mpc$). This relation measures an integrated quantity that correctly reproduces global properties.

In *section 8* we perform the multifractal analysis of the sample, by including also the luminosities of galaxies. Such an analysis allows us to study the correlation between galaxian space locations and luminosities. Indeed, the multifractal behavior shows the self-similar nature of the whole matter distribution. This properties imply a number of interesting consequences and we refer the reader to Baryshev et al.(1994), Sylos Labini & Pietronero (1995) for a more detailed discussion. Even in this case we find results that are in complete agreement with those obtained with the other methods discussed before, and with those of CP92.

We compare our results with those of other authors in *section 9* and finally we present our conclusions. The fact that the distribution of visible matter is fractal up to the present observational limits is fully consistent with the Cosmological Principle (namely that there are no preferential points in the Universe) since all points on a fractal structure are statistically equivalent. In fact, a fractal distribution is locally isotropic and non-analytic, so that it is also non-homogenous. The distinction between homogeneity and local isotropy is discussed in detail by Sylos Labini (1994) and Baryshev et al. (1994).

We stress that only in the case of an homogenous sample the amplitudes of correlation acquire a physical meaning. In the opposite case and anyhow for the range of scales in which the structure is self-similar (even if homogeneity is eventually achieved at large scale) it is necessary to change the theoretical language and perspective and adopt the one that is appropriate for self-similar and non-analytical structures (Pietronero & Tosatti, 1986; Baryshev et al., 1994).

## 2   Description of the data and subsamples

The *Perseus-Pisces* redshift survey collects the positions and the redshifts for the galaxies in the region $22^h < \alpha < 4^h$ and $0° < \delta < 45°$ (Giovanelli, & Haynes 1988). The survey consists mainly of highly accurate *21 cm* H I line redshifts. The radio data are complemented with optical observations of early-type galaxies carried out at the 2.4 m telescope of the MacGraw-Hill Observatory, plus a number of redshifts provided by J.Huchra and other smaller sources in the public domain. The catalog used comprises those redshifts obtained before 1991 December, for a total of 5183 galaxies. Among them, 3854 have Zwicky magnitudes of 15.7 or brighter. From this sample, we have excluded the data in the region more affected by extinction; hence we have considered the data only in the ranges $[22^h < \alpha < 3^h 10^m]$ and $[0° < \delta < 42°30']$.

The measured velocities of the galaxies have been expressed in the preferred frame of the Cosmic Microwave Background Radiation (CMBR), i.e. the heliocentric velocities of the galaxies have been corrected for the solar motion with respect to the CMBR, according with the formula:

$$\vec{v} = \vec{v}_m + 316 cos\theta \ \ kms^{-1} \qquad (1)$$

where $\vec{v}$ is the corrected velocity, $\vec{v}_m$ is the observed velocity and $\theta$ is the angle between the observed velocity and the direction of the CMBR dipole anisotropy ($\alpha = 169.5°$ and $\delta = -7.5°$). From these corrected velocities, we have calculated the comoving distances $r(z)$, with $q_0 = 0.5$:

$$r(z) = 6000 \left(1 - \frac{1}{\sqrt{(1+z)}}\right) h^{-1} Mpc \qquad (2)$$

We have studied galaxies with corrected velocity in the range $0 - 13,000$ $Kmsec^{-1}$. The apparent magnitudes of galaxies have been corrected for the extinction, using the absorption maps produced by Burstein & Heiles (Giovanelli et al., 1986). The final sample, with apparent magnitude less than 15.5, contains $N = 3301$ galaxies (that we call hereafter PP 15.5). With these data, we have produced some VL subsamples whose characteristic are reported in *Table 1*.

## 3   Essential properties of fractal structures

In this section we mention the essential properties of fractal structures because they will be necessary for the correct interpretation of the statistical analysis. However in no way these properties are assumed or used in the analysis itself. A fractal consists of a system in which more and more structures appear at smaller and smaller scales and the structures at small scales are similar to the one at large scales. Starting from a point occupied by an object, we count how many objects are present within a volume characterized by a certain length scale in order to establish a generalized "mass-length" relation from which one can define the fractal dimension. We can then write a relation between $N$ ("mass") and $r$ ("length") of type (Mandelbrot, 1982)

$$N(r) = B \cdot r^D \qquad (3)$$

where the fractal dimension is $D$ and the prefactor $B$ is instead related to the lower cut-offs . It should be noted that Eq.?? corresponds to a smooth convolution of a strongly fluctuating function. Therefore a fractal structure is always connected with large fluctuations and clustering at all scales. From Eq.?? we can readily compute the average density $<n>$ for a sample of radius $R_s$ which contains a portion of the fractal structure. The sample volume is assumed to be a sphere ( $V(R_s) = (4/3)\pi R_s^3$) and therefore

$$<n> = \frac{N(R_s)}{V(R_s)} = \frac{3}{4\pi} B R_s^{-(3-D)} \qquad (4)$$

From Eq.?? we see that the average density is not a meaningful concept in a fractal because it depends explicitly on the sample size $R_s$. We can also see that for $R_s \to \infty$ the average density $<n> \to 0$; therefore a fractal structure is asymptotically dominated by voids. We can define the conditional density from an occupied point as:

$$\Gamma(r) = S(r)^{-1}\frac{dN(r)}{dr} = \frac{D}{4\pi} B r^{-(3-D)} \qquad (5)$$

where $S(r)$ is the area of a spherical shell of radius $r$. Usually the exponent that defines the decay of the conditional density $(3-D)$ is called the codimension and it corresponds to the exponent $\gamma$ of the galaxy distribution.

We see therefore that the average density $<n>$ is not a well defined quantity, while the conditional average density, as given by Eq.??, is well defined in terms of its exponent, the fractal dimension. The amplitude of this function essentially refers to the unit of measures given by the lower cut-offs, but it has no particular physical meaning because this is not an intrinsic quantity.

## 4 $\Gamma(r)$ analysis

Fractal distributions are characterized by long-range power-law correlations. The correlation function (hereafter CF) of such a structure is described by (CP92)

$$G(r) = <n(r)n(0)> \approx r^{-\gamma} \qquad (6)$$

where the exponent $\gamma$ is the codimension (Eq.??) ($\gamma = 3-D$). If the sample is homogenous, $G(r) \approx <n>^2$ and hence it is constant. Therefore this CF is the appropriate statistical tool to study fractal versus homogeneity properties. For a more complete discussion we refer the reader to CP92.

To normalize the CF to the size of the sample under analysis we use, following CP92:

$$\Gamma(r) = \frac{<n(r)n(0)>}{<n>} = \frac{G(r)}{<n>} \qquad (7)$$

where $<n>$ is the average density of the sample. We stress that this normalization does not introduce any bias even if the average density is sample-depth dependent, as in the case of fractal distributions (Eq.??), because it represents only an overall normalizing factor. The CF of Eq.?? can be computed by the following expression

$$\Gamma(r) = \frac{1}{N} \sum_{i=1}^{N} \frac{1}{4\pi r^2 \Delta r} \int_{r}^{r+\Delta r} n(\vec{r}_i + \vec{r'})d\vec{r'} \qquad (8)$$

This function measures the average density at distance $r$ from an occupied point and it is called the *conditional density* (CP92). It is also very useful to use the conditional average density

$$\Gamma^*(r) = \frac{3}{4\pi r^3} \int_{0}^{r} 4\pi r'^2 \Gamma(r')dr' \qquad (9)$$

This function would produce an artificial smoothing of rapidly varying fluctuations, but it correctly reproduces global properties (CP92).

We have studied the behavior of $\Gamma(r)$ and $\Gamma^*(r)$ in the subsamples of *Table 1*. The results are shown in Fig.1. A well defined power law behavior is detected up to the sample limit without any tendency towards homogenization. The codimension is, with very good accuracy

$$\gamma = 3 - D \approx 1 \qquad (10)$$

so that $D \approx 2$ up to the sample limit. Hence the PP15.5 redshift survey shows well defined fractal properties up to the effective depth $R_s \approx 30h^{-1}Mpc$. It has consistent statistical properties and hence it is a *statistically fair* sample. Of course it is not an homogenous sample.

The fractal dimension $D \approx 2$, observed in this sample is somewhat larger than the value obtained for CfA that is $D \approx 1.5$. This could be due to the fact CfA1 has a reduced number of points. The value $D \approx 2$ is in agreement with the analysis of Guzzo et al. (1991) and Amendola & Sylos Labini (1995), and it is also consistent with CfA2 (Park et al., 1994; Sylos Labini & Amendola, 1995) and ESP (Pietronero & Sylos Labini, 1994; Sylos Labini et al., 1995) catalogues.

As discussed in CP92, we have limited our analysis to an effective depth $R_s$ that is the radius of the maximum sphere contained in the sample volume.

In such a way that we eliminate from the statistics the points for which a sphere of radius $r$ is not fully included within the sample boundaries. Hence we do not make use of any weighting scheme with the advantage that we do not make any assumption in the treatment of the boundaries conditions. Of course in doing this, we have a smaller number of points and we stop our analysis at a smaller depth than that of other authors (Guzzo et al., 1991), with the advantage, however, of not introducing any a priori hypothesis. We discuss this point more carefully in *section 9*.

# 5 $\xi(r)$ analysis

CP92 clarify some crucial points of the standard CF analysis, and in particular they discuss the meaning of the so-called "*correlation length*" $r_0$ found with the standard approach (Davis & Peebles, 1983) and defined by the relation:

$$\xi(r_0) = 1 \qquad (11)$$

where

$$\xi(r) = \frac{<n(\vec{r_0})n(\vec{r_0}+\vec{r})>}{<n>^2} - 1 \qquad (12)$$

is the two point correlation function used in the standard analysis. The basic point that CP92 stressed, is that the mean density used in the normalization of $\xi(r)$ is not a well defined quantity in the case of self-similar distribution and it is a direct function of the sample size. Hence only in the case that the homogeneity has been reached well within the sample limits the $\xi(r)$-analysis is meaningful, otherwise the a priori assumption of homogeneity is incorrect and the characteristic lengths, like $r_0$, became spurious.

For example, following CP92 the expression of the $\xi(r)$ in the case of fractal distributions is:

$$\xi(r) = ((3-\gamma)/3)(r/R_s)^{-\gamma} - 1 \qquad (13)$$

where $R_s$ is the depth of the spherical volume where one computes the average density from Eq.??. From Eq.?? it follows that i.) the so-called correlation length $r_0$ (defined as $\xi(r_0) = 1$) is a linear function of the sample size $R_s$

$$r_0 = ((3-\gamma)/6)^{\frac{1}{\gamma}} R_s \qquad (14)$$

and hence it is a spurious quantity without physical meaning but it is simply related to the sample finite size.

ii.) $\xi(r)$ is power law only for (from Eq.(6))

$$((3-\gamma)/3)(r/R_s)^{-\gamma} >> 1 \qquad (15)$$

hence for $r \lesssim r_0$: for larger distances there is a clear deviation from a power law behavior due to the definition of $\xi(r)$. This deviation, however, is just due to the finite size of the observational sample and does not correspond to any real change of the correlation properties. The analysis performed with $\xi(r)$ is therefore mathematically inconsistent, if a clear cut-off towards homogeneity has not been reached, because it gives an information that is not related with real physical features of the sample but to the geometry of the sample itself.

We have studied the $\xi(r)$ in the VL subsamples of *Table 1*. We found that for $r \lesssim r_0$ $\gamma \approx 1$ as shown in Fig.2. The amplitude of $\xi(r)$ is sample depth dependent: in Fig.3 the behavior of $r_0$ is plotted as a function of the sample depth $R_s$. One can see that the experimental data are very well fitted by Eq.??. This analysis is in agreement with that of CP92 and with the analysis done by the CF $\Gamma(r)$ discussed previously. The so-called "correlation length" $r_0$ has therefore no physical meaning but it only represents a fraction of the sample size in the sense of Eq.??. Several authors (Da Costa et al, 1988, Park et al. 1994) have discussed the shift of $r_0$ as an effect of luminosity segregation. We show in *section 6* that this interpretation is not supported by any experimental evidence and that the shift of $r_0$ is naturally described by the fractal nature of galaxy distribution in this sample.

## 6   About the luminosity segregation

A possible explanation of the shift of $r_0$ is based on the luminosity segregation effect (Da Costa et al., 1988; Park et al., 1994). We briefly illustrate this approach. The fact that the giant galaxies are more clustered than the dwarf ones, i.e. that they are located in the peaks of the density field, has given rise to the proposition that larger objects may correlate up to larger length scales and that the amplitude of the $\xi(r)$ is larger for giants than for dwarfs one. The deeper VL subsamples contain galaxies that are in average brighter than those in the VL subsamples with smaller depths. As the brighter galaxies should have a larger correlation length the behavior found in Fig.3 could be related, at least partially, with the phenomenon of luminosity segregation.

To show that this is not the case for the PP15.5 survey and that, on the contrary, the shift of $r_0$ is simply due to the fractal nature of the galaxy distribution, we have performed the following test. We consider a sample of galaxies with *apparent magnitude* less than 14.5 (hereafter PP14.5). In Fig.4. it is shown the absolute-magnitude versus distance diagram for the two catalogs: the whole PP15.5 and PP14.5. It is evident that the VL subsample with the same absolute magnitude limit $M_{lim}$ have different limiting depth $d_{lim}$ for the two catalogs according to the formula

$$d_{lim} = 10^{0.2 \cdot (m_{lim} - M_{lim} - 25)} \tag{16}$$

where $m_{lim}$ is 15.5 or 14.5. Then we construct some VL subsamples for the PP14.5 catalog whose characteristics are reported in *Table 2*. For these VL subsamples we have done the same analysis as for the whole catalog PP15.5. For example the subsample VL75(b) (we refer with (b) to the 14.5 sample) has the same absolute magnitude limit of the sample VL120 of PP15.5. Hence these two subsamples contain galaxies with the same average absolute magnitude. If the shift of $r_0$ is due to the luminosity segregation effect we should not find any difference for $r_0$ in these two subsamples. As shown in Fig.5. this is not the case. In fact, for VL75(b) we find that $r_0 \approx 8.5 h^{-1} Mpc$ while for VL120 $r_0 \approx 11 h^{-1} Mpc$.

On the contrary, if we consider the sample depth dependence of $r_0$ we find that for VL75(b) $R_s = 22.6 h^{-1} Mpc$ that is the same limiting depth of the subsample VL80 for which $r_0 \approx 8 h^{-1} Mpc$ as shown in Fig.5. The behavior of $r_0$ with the sample depth in the PP14.5 sample, is well fitted by Eq.?? as for the whole catalog PP15.5 (Fig.5).

Our conclusion is that this analysis shows unambiguously that the luminosity segregation effect cannot be the cause of the shift of $r_0$ and that, on the contrary, the linear dependence of $r_0$ is naturally described by the fractal nature of the galaxy distribution.

We are going to see that the observation that the giant galaxies are more clustered than the dwarf ones, i.e. that the massive elliptical galaxies lie in the peaks of the density field, is a consequence of the self-similar behavior of the whole matter distribution (*section 8*). The increasing of the correlation length of the $\xi(r)$ has nothing to do with this effect (CP92, Baryshev et al., 1994).

# 7 Number-distance and number-counts relations

We study now the behavior of two classical relations for which there are definite predictions in the case of a homogenous distribution (Sandage, 1988) as well as for fractal distributions (Baryshev et al., 1994). At these small redshifts we do not consider the effect of galaxy and space-time evolution. We have computed the *number-distance relation*: the behavior of the $N(r)$ relation, the integrated number of galaxies with distance less than $r$, for an homogenous Universe, at small distance, is simply:

$$N(r) \sim r^3 \qquad (17)$$

while if the sample has fractal nature the exponent of Eq.?? is equal to the fractal dimension $D < 3$ (see Eq.??). We have studied Eq.?? in the VL subsamples of *Table 1*. The results are shown in Fig.6. From these data we can measure the fractal dimension that is $D \approx 2$ up to the sample limiting depth ($R_s = 140 - 150 \ h^{-1} Mpc$). This result is in perfect agreement with the fractal dimension measured by the CF $\Gamma(r)$. In any case we stress that the right behavior can be found only if there are enough points.

The different normalization in the various VL samples in Fig.6 is due to the different absolute magnitude limit ($M_{lim}$) that define each VL subsample. To normalize the behavior of $N(r)$ we divide it for a luminosity factor given by:

$$\Phi(M_{lim}) = \int_{-\infty}^{M_{lim}} \phi(M) dM \qquad (18)$$

where $\phi(M)$ is the Schechter luminosity function. As parameters of $\phi(M)$ we take $\alpha = -1.2$ and $M_* = -19.5$ according to Da Costa et al. (1994). The amplitude of the luminosity function does not enter in the normalization between different VL subsamples.

In Fig.7 we show the behavior of the average density

$$n(r) = \frac{N(r)}{V(r)} \sim r^{-(3-D)} \quad (19)$$

for several VL samples, normalized to the luminosity factor $\Phi(M_{lim})$ of Eq.**??**. From this figure we can see that there is a nice agreement of the various VL subsamples, in the scaling region.

We stress that this method is more subjected to statistical fluctuations than the CF analysis because in the latter case one performs an average from *each point* of the sample. On the contrary, in this case, we compute $N(r)$ only from *one point* but with the advantage to have more galaxies and a larger volume, because we are not limited by the request to have a spherical sample inside the volume of the survey as we have done for the CF analysis (see *section 4*), in order to do not make any assumption on the data analysis. Therefore at small scales $N(r)$ is not well defined, because of too few points. However $N(r)$ allows us to extend the analysis to lengths scales that cannot be reached by the full correlation analysis and it measures an integrated quantity that correctly reproduces global properties. Moreover we stress that Eq.**??** (and Eq.**??**) for a fractal distribution corresponds to a convolution of strongly fluctuating quantities, while if the distribution is homogenous the fluctuations are intrinsically of small amplitude. At $50h^{-1}Mpc$ there is a very large structure known as the Perseus-Pisces supercluster (Haynes & Giovanelli, 1988) that locally affects the behavior of $N(r)$ and it is the cause of the initial fast increase.

In any case these results show that, if the number of galaxies is large enough, we can recover the correct fractal dimension by means of the number-distance analysis. In fact, we find the same exponent that comes out from the $\Gamma(r)$ analysis.

Therefore our conclusion is that the sample shows fractal nature with $D \approx 2$ up to its *limiting depth* that is $\approx 120 - 140h^{-1}Mpc$.

The number-counts relation $N(m)$ gives the number of galaxies with apparent magnitude lower than $m$ versus $m$. It is simple to show that if the sample is fractal with dimension $D$ we have:

$$\log(N(m)) \sim \alpha m \quad (20)$$

with $\alpha = D/5$ (Peebles, 1993): if $D = 3$ $\alpha = 0.6$. From the previous relation it seems that, by knowing the luminosity properties of galaxy distribution as expressed by the count-magnitude relation (Eq.**??**), it is possible to reconstruct the fractal exponent of the space distribution. However this is correct only if there are no correlations between space locations and luminosities of galaxies. If we do not *assume* this property we have to study the $N(m)$ relation in VL subsamples. Therefore we have performed such an analysis for the VL subsamples of *Table 1*. As shown in Fig.8 we find that

$$\alpha \approx 0.4 = \frac{D}{5} \qquad (21)$$

with $D = 2$. Hence we have obtained the correct exponent consistent with to Eq.??. We have computed also $N(m)$ for the whole magnitude limited PP catalog. In this case we find that $\alpha \approx 0.6$ (Fig.9.). Clearly it would be naive to argue that this corresponds to homogeneity. Namely we know from the $\Gamma(r)$ and $N(r)$ analyses that the sample has fractal properties. The selection effects in the magnitude limited survey affect the correct behavior of $N(m)$ and increase the value of $\alpha$ from $D/5 \approx 0.4$, that we find only in VL samples, to 0.6. This is an important results that should been taken into account in the whole galaxy counts analysis (Baryshev et al., 1994).

Our conclusion is, that, this test is quite delicate. If properly performed, it is in agreement with the $\Gamma(r)$ and $N(r)$ analyses. Moreover we found that the Malmquist bias, present in magnitude limited (hereafter ML) samples, affects the behavior of $N(m)$ so that one should always use VL rather than ML subsample to study this behavior. Finally we have an indication that the space locations of galaxies are correlated with their luminosities. This correlation can be studied by means of the multifractal analysis that we show in *section 8*.

# 8   Multifractal analysis

The multifractal picture is a refinement and generalization of the fractal properties (Paladin & Vulpiani, 1987; Benzi et al., 1984, CP92). In no way it is in contrast with it and the fractal study of a problem is just a subset of the full multifractal analysis In the simple fractal case one needs only an exponent to characterize the scaling properties of the system. The situation can be however more complex and the scaling properties can be different for different regions of the system. In this case one has to introduce a continuous set of fractal indexes to characterize the system (the multifractal spectrum). One refers to this case with the term "multifractality" (hereafter MF). This more general concept is necessary in the case of a self-similar measure (i.e. if the galaxy masses are included). In this case it also acquires an important physical meaning. It is rather pointless instead to consider the eventual multifractal properties of the support of the measure (i.e. the galaxy number density) (Martinez & Jones, 1990) and we will not consider the question.

The discussion that we have presented in the previous sections, was meant to distinguish between homogeneity and scale invariant properties and, for this purpose, it is perfectly appropriate even if the galaxy distribution would be multifractal. In this case the correlation functions we have considered would correspond to a single exponent of the multifractal spectrum, but the issue of homogeneity versus scale invariance (fractal or multifractal) remains exactly the same.

The distribution of visible matter is described, in a certain sample of depth $R_s$, by the density function:

$$\rho(\vec{r}) = \sum_{i=1}^{N} m_i \delta(\vec{r} - \vec{r_i}) \qquad (22)$$

where $m_i$ is the mass of the $i$-th galaxy. This distribution corresponds to a measure defined on the set of points which have the correlation properties described by Eq.??. It is possible to define the normalized density function:

$$\mu(\vec{r}) = \sum_{i=1}^{N} \mu_i \delta(\vec{r} - \vec{r_i}) \qquad (23)$$

with $\mu_i = m_i/M_T$ and $M_T = \sum_{i=1}^{N} m_i$. The quantity $\mu(\vec{r})$ is dimensionless. Suppose that the total volume of the sample consists of a *3*-dimensional cube of size $L$. We divide this volume into boxes of linear size $l$. We label each box by the index $i$ and construct for each box the function:

$$\mu_i(\epsilon) = \int_{i-thbox} \mu(r) dr \qquad (24)$$

where $\epsilon = l/L$. In the case of a MF distribution if in the $i$-th box there is a singularity of type $\alpha$ then in the limit $\epsilon \to 0$, the measure goes as:

$$\mu_i(\epsilon) \sim \epsilon^{-\alpha(\vec{x})} \qquad (25)$$

In Eq.?? the exponent $\alpha(\vec{x})$ (a sort of local fractal dimension) fluctuates widely with the position $\vec{x}$. For an homogeneous mass distribution, with a uniform density, $\alpha = 3$, while for a simple fractal with dimension $D$, $\alpha = D$. In general we will found several boxes with a measure that scales with the same exponent $\alpha$. These boxes form a fractal subset with dimension $f(\alpha)$. Hence the number of boxes that have a measure that scale with exponent in the range $[\alpha, \alpha + d\alpha]$ vary with $\epsilon$ as:

$$N(\alpha, \epsilon) d\alpha \sim \epsilon^{-f(\alpha)} d\alpha \qquad (26)$$

The function $f(\alpha)$ is usually (Paladin & Vulpiani, 1987) a single humped function with the maximum at $max_\alpha f(\alpha) = D(0)$, where $D(0)$ is the dimension